\documentclass[twoside]{article}

\usepackage[preprint]{arxiv}

\usepackage[round]{natbib}

\usepackage[utf8]{inputenc}
\usepackage{enumerate}
\usepackage{amsmath}
\usepackage{amssymb}
\usepackage{bbm}
\usepackage{hyperref}
\hypersetup{
    colorlinks,
    linkcolor={red!50!black},
    citecolor={blue!50!black},
    urlcolor={blue!80!black}
}

\usepackage{tikz}
\usepackage{xcolor}
\usepackage{booktabs}
\usepackage{multirow}
\usepackage{multicol}
\usetikzlibrary{arrows,automata}
\newdimen\arrowsize
\pgfarrowsdeclare{arcsq}{arcsq}
{
  \arrowsize=0.2pt
  \advance\arrowsize by .5\pgflinewidth
  \pgfarrowsleftextend{-4\arrowsize-.5\pgflinewidth}
  \pgfarrowsrightextend{.5\pgflinewidth}
}
{
  \arrowsize=1.5pt
  \advance\arrowsize by .5\pgflinewidth
  \pgfsetdash{}{0pt} 
  \pgfsetroundjoin   
  \pgfsetroundcap    
  \pgfpathmoveto{\pgfpoint{0\arrowsize}{0\arrowsize}}
  \pgfpatharc{-90}{-140}{4\arrowsize}
  \pgfusepathqstroke
  \pgfpathmoveto{\pgfpointorigin}
  \pgfpatharc{90}{140}{4\arrowsize}
  \pgfusepathqstroke
}

\definecolor{aliceblue}{rgb}{0.94, 0.97, 1.0}
\usepackage[color=aliceblue]{todonotes}
\usepackage[most]{tcolorbox}
\usepackage{subcaption}

\newcommand{\indep}{\perp \!\!\! \perp}

\newtcolorbox{aside}{%
    enhanced,%
    attach boxed title to top center={yshift=-3mm,yshifttext=-1mm},%
    colback=blue!5!white,tlmgr -check%
    colframe=blue!75!black,%
    fonttitle=\bfseries,%
}

\begin{document}

\runningtitle{Label Shift Estimators for Non-Ignorable Missing Data}
\runningauthor{Andrew C.~Miller, Joseph Futoma}

\twocolumn[
\aistatstitle{Label Shift Estimators for Non-Ignorable Missing Data}
\aistatsauthor{ Andrew C.~Miller \\ \texttt{acmiller@apple.com} \And Joseph Futoma \\ \texttt{jfutoma@apple.com}}
\aistatsaddress{ Apple \And Apple }

]

\begin{abstract}

We consider the problem of estimating the mean of a random variable $Y$ subject to non-ignorable missingness, i.e., where the missingness mechanism depends on $Y$.  We connect the auxiliary proxy variable framework for non-ignorable missingness \citep{west2013non} to the label shift setting \citep{saerens2002adjusting}. Exploiting this connection, we construct an estimator for non-ignorable missing data that uses high-dimensional covariates (or proxies) without the need for a generative model.  
In synthetic and semi-synthetic experiments, we study the behavior of the proposed estimator, comparing it to commonly used ignorable estimators in both well-specified and misspecified settings. 
Additionally, we develop a score to assess how consistent the data are with the label shift assumption. 
We use our approach to estimate disease prevalence using a large health survey, comparing ignorable and non-ignorable approaches.  We show that failing to account for non-ignorable missingness can have profound consequences on conclusions drawn from non-representative samples.
\end{abstract}


\vspace{-1.5em}
\section{Introduction}

\textbf{}\begin{figure}[t!]
    \centering
    \includegraphics[width=0.82\columnwidth]{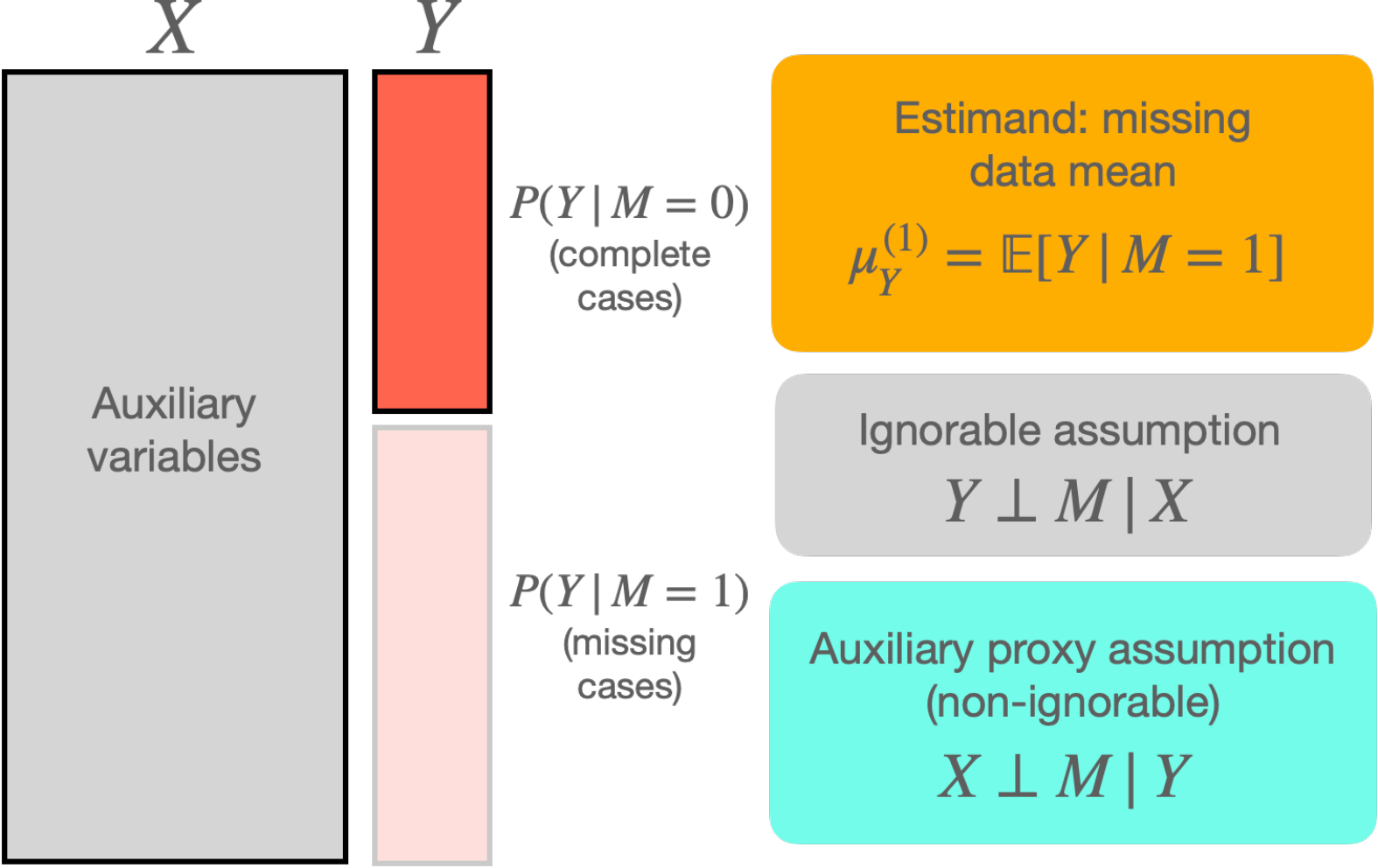}
    \caption{\textbf{Overview}.  The target of our estimation procedure is the expected value of $Y$ for the missing cases, $\mu_Y^{(1)}$.  A set of fully observed covariates $X$ are used to form the estimate.  Under the common ignorability assumption, observing $X$ renders $Y$ and $M$ independent.  We develop estimators for the non-ignorable setting, using a different assumption from the auxiliary proxy variable framework to aid identification.}
    \label{fig:overview}
    \vspace{-.5em}
\end{figure}

\vspace{-2.2em}
Missing data can frustrate the construction of valid statistical inferences. 
Study data --- particularly from surveys or health records --- often feature missingness of unknown mechanism.
The typical goal of statistical inference is to characterize some aspect of the full population, e.g., the prevalence of a disease over both observed \emph{and} unobserved subjects.
As the observed data are often straightforward to summarize, characterizing the unobserved data is the main challenge.
In this work, we focus on estimating the mean of a random variable subject to missingness.
This task appears in many different applications --- e.g., estimating disease prevalence in a population with selectively observed labels, or measuring the expected performance of a prediction model deployed on a population that differs from the study data used for training. 


Without introducing assumptions, however, we are generally unable to estimate properties of the missing data, as they may not be \emph{identified} by the observed data alone \citep{rubin1976inference}.
Even with infinite data we would be no more certain about the true quantity of interest, as there may be different values that are equally compatible with the observed data.
Well-specified assumptions enable the estimation of missing data parameters from observed data alone.

Perhaps the simplest missing data approach is to restrict analysis to subjects where all data are observed, i.e., \emph{complete cases}.
Though seemingly innocuous, such an analysis implicitly introduces a strong assumption --- that missing units are statistically identical to observed units. 
However, when missing units systematically differ from observed units (e.g., selection bias), a complete case analysis can produce misleading and invalid inferences. 

Another common, but far less stringent, assumption is \emph{ignorability}  ---  sometimes referred to as ``no unmeasured confounding'' --- which assumes that fully observed covariates can
 explain the missing mechanism \citep{rubin1976inference}. 
Despite its ubiquity in missing data applications, many real-world scenarios violate ignorability, which can also lead to invalid inferences.

For such \emph{non-ignorable} settings, less common strategies exist that can produce valid inferences, leveraging different assumptions.
One such approach, which we extend in this work, is the auxiliary proxy variable framework \citep{west2013non}.
However, a major limitation is that it requires modeling a high-dimensional conditional distribution, which can be intractable or impose rigid distributional constraints. 

In this paper, we identify a key connection between the proxy variable approach and the label shift problem \citep{saerens2002adjusting}, a commonly analyzed setting within the broader area of robustness and distribution shift in machine learning.
We show that label shift is intimately related to the proxy variable approach, allowing us to adapt machine learning estimators for label shift to address the non-ignorable missing data problem. 
Leveraging this insight, we develop a measure of how consistent the observed data are with the common assumption underlying both the auxiliary proxy variable and label shift paradigms.
This measure depends on the \emph{propensity score}, a common tool in the ignorable missing data setting that we show is also useful in the non-ignorable and label shift settings.

Specifically, in this work we:
\vspace{-.5em}
\begin{itemize} \setlength\itemsep{-.1em}
    \item connect the non-ignorable framework of \citet{west2013non} to the label shift problem \citep{saerens2002adjusting}, and develop an estimator that can flexibly incorporate high-dimensional covariates;
    \item devise the \emph{propensity coherence score}, a measure of how consistent the observed data are with the auxiliary proxy (and label shift) assumption, and show that it tracks with estimation quality; 
    \item conduct synthetic and semi--synthetic simulation studies, comparing ignorable, non-ignorable, and high-dimensional proxy variable estimators under various missing data regimes, with and without misspecification;
    \item detail a case study comparing estimators applied to natural missingess in a large nationally representative health survey.
\end{itemize}
\vspace{-.5em}
Our empirical study shows that naively assuming ignorability may produce misleading inferences, and directly addressing non-ignorability can yield valid inferences in certain situations.  
Unsurprisingly, we find that no single approach dominates, and careful sensitivity analyses are needed in practice. 
Our analysis also highlights the importance of calibrating predictors, as this can have a profound impact on the final estimates, particularly in high-dimensional settings.

\vspace{-0.5em}
\section{Method}


\subsection{Background and notation}
\label{sec:background}

We consider scenarios with a binary target variable $Y \in \{0, 1\}$ and a missingness indicator $M \in \{0, 1\}$, where $M=0$ denotes that $Y$ is observed.
Additionally, we observe a covariate vector $X$ of auxiliary information for both the $M=1$ and $M=0$ cases. 
The goal is to estimate the mean of the missing $Y$, $\mu_Y^{(1)} \triangleq \mathbb{E}[Y \,|\, M=1]$ --- see Figure~\ref{fig:overview}. 
This can then be combined with the estimated mean of the observed values, $\mu_Y^{(0)} \triangleq \mathbb{E}[Y \,|\, M=0]$, to estimate the overall mean for $Y$ as $\mathbb{E}[Y] = \sum_{m \in \{0,1\}} P(M=m) \mu_Y^{(m)}$.\footnote{
Note that we will occasionally switch to lowercase letters as shorthand for probability density or mass functions, as relevant (e.g. $P(Y=y) = p(y)$).}

\vspace{-.5em}
\paragraph{Complete case analysis}
The most naive approach to estimating the mean of the missing $Y$ values is to simply use an estimate of the mean of the observed $Y$ values.  
That is, assume that $\mu_Y^{(1)} = \mu_Y^{(0)}$, and then assume that any inferences about $\mu_Y^{(0)}$ will directly translate to estimating $\mu_Y^{(1)}$. 
This is referred to as \emph{complete case analysis}, as we only utilize data from the complete or observed cases. 
It makes the strong assumption that the missingness is completely at random, i.e. that $Y \indep M$ and hence all observations share the same missingness probability $\pi = P(M=1)$, which is rarely satisfied in practice.

\vspace{-.5em}
\paragraph{Ignorable missing data}
The prevailing approach to missing data assumes \emph{ignorability} --- that there is enough information in the covariates $X$ to render the target and the missing indicator independent, i.e., $Y \indep M \,|\, X$. 
This is often informally referred to as ``no hidden confounding.'' 
For instance, $X$ might denote access to healthcare, which can causally influence both the presence of disease and whether it has been formally diagnosed yet.  
Marginally, there may be dependence between $Y$ and $M$ (e.g., the unobserved $M=1$ population might have a higher prevalence of disease), but ignorability assumes that observing $X$ renders them independent, allowing for statistical estimation of the mean of $Y$ in the unobserved population on the basis of observed data alone.
When this assumption is valid, a variety of estimators for $\mu_Y^{(1)}$ can be deployed, including the inverse probability weighted estimator (IPWE) \citep{robins1994estimation}, the direct regression estimator \citep{kang2007demystifying}, and doubly robust estimators that combine them  \citep{cao2009improving}. 
However, when ignorability does not hold (and in many real situations, it probably does not), these estimators can be inaccurate. 

\vspace{-.5em}
\paragraph{Non-ignorable missing data}
In non-ignorable settings, a factorization of the joint distribution of $(X, Y,M)$ typically guides assumptions and estimation.
There exist multiple factorizations \citep{franks2020nonstandard}, with the two main classes being the \emph{pattern mixture model} \citep{little1994class, west2013non} and the \emph{selection model} \citep{heckman1976}.
The selection model factorizes this as $P(M,X,Y) = P(M|X,Y) P(X,Y)$, and separately models the selection or missing data mechanism from the model for $X$ and $Y$. 
If the selection mechanism is ignorable, this simplifies to $P(M,X,Y) = P(M|X) P(X,Y)$.
However, our primary interest is in the pattern mixture model, which factorizes as $P(M,X,Y) = P(M) P(Y,X|M)$, defining the marginal distribution $P(X,Y)$ as a mixture of different distributions for the observed and missing populations.  Furthermore, one way to write out this data generating process makes explicit which parameters can be readily identified:
\begin{align}
    M &\sim \texttt{Bern}(\pi) \,, \\
    Y | M &\sim \texttt{Bern}(\mu_Y^{(M)}) \,, \\
    X | Y, M &\sim P(X \,|\, Y, \theta^{(M)}) \,. 
\end{align}
It is clear that $\mu_Y^{(0)}$ is simply the mean of the observed $Y$ variables, and $\theta^{(0)}$ parameterizes the conditional of $X$ given $Y$ when the $Y$ are observed.  
However, $\theta^{(1)}$ and $\mu_Y^{(1)}$ --- our estimand --- in general cannot be identified by observed data alone.
Thus, we need to make an additional assumption about similarities or invariances across levels of missingness, $M=0$ and $M=1$, in order to render $\mu_Y^{(1)}$ identifiable.


\vspace{-.5em}
\paragraph{Auxiliary variables for pattern mixtures}
One way to identify $\mu_Y^{(1)}$ invokes a specific conditional independence assumption about the auxiliary variables, or covariates, $X$ \citep{west2013non}.
These estimators apply when the missingness (e.g., having had a diagnostic test ordered) depends on the true $Y$ value but not on the other available covariates $X$ after conditioning on $Y$, i.e., $X \indep M \,|\, Y$. 
Intuitively, if we know $Y$, then $X$ would give us no additional information about $M$.  
As a concrete example, let $Y$ denote the result of a diabetes lab test, $X$ denote certain risk factors such as BMI or family history of diabetes, and $M$ (i.e., the presence of the diabetes test) denote the clinical suspicion of current diabetes. If we assume that $M$ is primarily influenced by findings observed in the clinic but not captured in $X$ --- e.g., a noisy version of $Y$ --- then you would not expect $X$ to influence $M$ once we know $Y$.
We refer to this as the \emph{stable proxy assumption}.
The stable proxy assumption implies that the conditional distribution {${P(X \,|\, Y, M=m)}$} is stable across both levels of $M$.  
This is quite different from the ignorability assumption, which implies stability in a different conditional distribution,  $P(Y \,|\, X, M=m)$.
Previous work with the stable proxy assumption developed estimators for continuous-valued $Y$ and scalar $X$, while assuming the parametric form of the (stable) conditional distribution $P(X \,|\, Y)$ to be Gaussian \citep{west2013non}.



\vspace{-.5em}
\paragraph{Label shift}
Forming predictions when the training (or source) distribution differs from the test (or target) distribution is a major challenge in statistics and machine learning.  
One approach to this problem exploits the \emph{label shift} assumption --- that the source distribution $P$  differs from the target distribution $Q$ only in the marginal of $Y$, i.e., $P(X \,|\, Y) = Q(X \,|\, Y)$ and $P(Y)\neq Q(Y)$.
Intuitively, such a situation may occur when $Y$ \emph{causes} $X$, e.g., a disease $Y$ causes symptoms $X$ via the same mechanism, but may occur at a different prevalence $Q(Y)$ in the target population.

Previous work has addressed the label shift (or shifting base rate) problem to adjust predictions at test time \citep{saerens2002adjusting, alexandari2020maximum}.  
One approach makes use of the \emph{scaled likelihood trick} \citep{renals1994connectionist}.
The key idea is that the likelihood of the generative model $p(x \,|\, y)$ can be estimated up to a constant factor using only a discriminative model $p(y \,|\, x)$, that is,
\begin{align}
    p(x \,|\, y) \propto p(y \,|\, x) \,/\, p(y) \,.
\end{align}
Adapting a learned prediction model $p(y \,|\, x)$ to $Q$, i.e., $q(y \,|\, x)$, involves a simple rescaling of the source model:
\vspace{-1em}
\begin{align}
    q(y \,|\, x) 
    &\propto p(x \,|\, y) q(y)\,, && \text{\color{gray}{label shift assumption}} \\
    &= p(y \,|\, x) \frac{q(y)}{p(y)} \,. && \text{ \color{gray}{adaptation }}
\end{align}
The label shift assumption only requires estimating the ratio of the prevalence of $Y$ from target to source, $w(y) \triangleq q(y) / p(y)$, to adapt the source predictor. 

Estimators in the label shift literature either estimate the ratio $w(y)$ directly, or form estimates of $p(y)$ and $q(y)$ separately.
Given only unlabeled samples from the target distribution, i.e., $X \sim Q(X)$, one effective strategy is to maximize the log-likelihood of the unlabeled data given estimates of $p(y \,|\, x)$ and $p(y)$, with respect to a single parameter $\pi_y = q(y)$.  
For a single $x$ from $Q$, this corresponds to
\begin{align}
    \ln q(x \,|\, \pi) &= \ln \left( \sum_{y} \pi_y \, p(x \,|\, y) \right) \\ 
    &= \ln \left( \sum_{y} \pi_y \frac{p(y \,|\, x)}{p(y)} \right) + \text{const.}\, , 
    \label{eq:likelihood}
\end{align}
where fixed estimates of $p(y \,|\,x)$ (i.e., a discriminative model) and $p(y)$ (i.e., the source base rate) can be plugged in.
Provided that the plug-in estimates are well-calibrated, expectation maximization (EM) is an effective and efficient way to optimize such an objective to form an estimate $\hat{\pi}_y$ \citep{alexandari2020maximum}.

\subsection{Connecting label shift and the auxiliary proxy variable framework}
\label{sec:connection}
Our first contribution is to make a simple connection explicit --- that the \emph{stable proxy assumption} deployed in non-ignorable missing data problems and the \emph{label shift assumption} in distribution shift problems are identical. 
We can view the non-ignorable missing data estimation problem as a type of label shift --- the source distribution $P$ corresponds to the complete cases $M=0$ where we observe $Y$, and the target distribution $Q$ corresponds to the cases $M=1$ where we are missing $Y$. 
In that sense, our estimand is simply an inferential byproduct of the label shift estimation problem, since 
for binary $Y$,
\begin{align}
    \mu_Y^{(1)} &= \mathbb{E}[Y \,|\, M=1] = \mathbb{E}_Q[Y] = \pi_1 \,.
\end{align}
The crucial assumption in the auxiliary proxy variable estimator is that the missingness $M$ is conditionally independent of covariates $X$ given $Y$.  Our key insight is that this directly implies that the class conditional distributions of $X$ given $Y$ are stable:
\begin{align*}
    X \indep M \,|\, Y &\implies P(X \,|\, Y, M=0) = P(X \,|\, Y, M=1) \, ,
\end{align*}
which is exactly the label shift assumption.
This equivalence allows us to import estimators from the label shift setting --- e.g., the calibration and EM procedure ~\citep{alexandari2020maximum} --- into the non-ignorable missing data setting.
The main difference is that our goal is no longer an adapted prediction model corresponding to the target $q(y \,|\, x)$, but instead the prevalence $q(y)$ --- typically a byproduct in the label shift problem.

\subsection{Proposed estimation routine}
\label{sec:algorithm}
A limitation of the auxiliary proxy variable framework of \citet{west2013non} is the need to specify a parametric distribution for $X$ given $Y$ (for both $M=0$ and $M=1$).
We can directly apply insights from the distribution shift literature to aid estimation in this setting. 
Indeed, the flexibility of the label shift estimator will allow us to relax distributional assumptions about $P(X \,|\, Y, \theta^{(m)})$, instead using machine learning techniques to form an estimate of $P(Y \,|\, X)$.  We propose the following procedure for estimating $\mu_Y^{(1)}$, following \citet{alexandari2020maximum}:
\vspace{-.5em} 
\begin{enumerate}[(i)] \setlength\itemsep{0em}
    \item split the observed cases where $M=0$ into a training set and a validation set for calibration, and add all the unobserved $M=1$ cases where we only have $X$ to the validation set;
    \item fit a machine learning model $\hat{y} = f(x) \approx p(y \,|\, x)$ to predict $Y$ on the training set of observed cases, and compute prediction scores $\hat{\mathcal{Y}}^{(0)}$ on the validation set of observed cases; 
    \item calibrate the scores $\hat{\mathcal{Y}}^{(0)}$ using a standard technique, such as Platt scaling \citep{platt1999probabilistic}, and compute calibrated scores on the missing validation cases, $\hat{\mathcal{Y}}^{(1)}$;
    \item using the calibrated scores on the missing validation cases $\hat{\mathcal{Y}}^{(1)}$, maximize the likelihood of their covariates $X$ with respect to $\mu_Y^{(1)}$ (Eq~\ref{eq:likelihood}), e.g., via EM \citep{saerens2002adjusting}, returning $\hat{\mu}_Y^{(1)}$.
\end{enumerate}
\vspace{-.5em}
In our experiments, we approximate confidence intervals with respect to the sampling distribution with bootstrap samples by repeating step (iv) for $B=1{,}000$ bootstrap replicates.

\vspace{-.5em}
\paragraph{Model calibration}
Machine learning models fit using empirical risk minimization can exhibit poor calibration \citep{guo2017calibration}. 
Crucially, the procedure detailed in the previous section requires the interpretation of $\hat{y} = p(y \,|\, x, m=0)$ as a calibrated probability in the observed distribution.  
Following the advice of \citet{alexandari2020maximum} we use Platt scaling \citep{platt1999probabilistic} to re-calibrate model scores
\begin{align}
p(y \,|\, x) \propto \exp\left(\frac{1}{T} \texttt{logit}(\hat{y}) + b\right) \, .
\end{align}
This is implemented as a logistic regression, fitting the two parameters $T$ and $b$ on the validation set. Other approaches such as isotonic regression are also possible \citep{niculescu2005predicting}.


\subsection{Propensity coherence score}
In general, the stable proxy (i.e., label shift) assumption may not exactly hold. 
In this case, assuming stability of the class conditional distributions may produce incorrect and overconfident estimates.
Though we cannot form a consistent estimate for the missing data mean in a completely general setting, we can devise a measure of how consistent the stable proxy assumption is with our observations. 
We do so by comparing two separate ways of estimating the propensity score, $P(M \,|\, X)$ --- one that uses the label shift assumption, and one that does not. 
When the label shift assumption is true, we can show that the two are mathematically equivalent (see Appendix~\ref{sec:consistency} for full derivation), and so the difference between the two methods should be small (due to finite sample errors and misspecification).
Hence, quantifying the difference between the estimated scores from the two approaches provides a notion of how divergent the observed data is from the label shift assumption.

The first approach is straightforward --- fit a calibrated prediction model for $P(M \,|\, X)$, i.e., the typical way propensities are estimated for ignorable missing data.
We let $P^{(\text{direct})}(M \,|\, X)$ denote our direct estimate of the propensity scores via a supervised model.

The second approach leverages the stable proxy assumption. 
We can rewrite $P(M \,|\, X)$ in terms of $\mu_Y^{(1)}$, $\mu_Y^{(0)}$, the discriminative model scores $P(Y \,|\, X, M=0)$, and the prevalence $P(M)$:
\begin{align}
    p(m | x) \propto \left(\sum_{y \in \{0,1\}} \frac{p(y \,|\, x, m=0)}{p(y \,|\, m=0)} p(y \,|\, m) \right) p(m) \, ,
\end{align}
all of which we already estimated.
We denote this second estimate of the propensities, achieved via the stable proxy assumption, by $P^{(\text{stable})}(M \,|\, X)$.
We quantify the difference between the two sets of estimated propensities, which we call the \emph{propensity coherence score}, with the symmetric KL divergence\footnote{We explored other methods to quantify the difference (e.g. the mean difference in log-likelihoods of the missingness indicators under each set of scores, the mean squared or mean absolute error between the scores or between the log of the scores, etc) and found they exhibited similar qualitative behavior to symmetric KL.}:
\begin{align}
    \delta = \mathbb{E}_X\left[ D_{KL} \left( P^{(\text{direct})}(M \,|\, X) \,||\, P^{(\text{stable})}(M \,|\, X) \right) \right] \, .
\end{align}
The coherence score supplies a measure of how consistent our observed data are with label shift (assuming we are confident that $P(M \,|\, X)$ and $P(Y \,|\, X, M=0)$ are reasonably well specified and calibrated).

\subsection{Related Work}
No references we are aware of establish an explicit connection between missing data estimators and label shift, both active areas of research in statistics and machine learning. 
Proxy pattern mixture analysis \citep{andridge2011proxy}, a related approach for incorporating high dimensional proxy variables, regresses $Y$ on a set of covariates $X$ to find the predictor that correlates most highly with $Y$, but still requires parametric assumptions for this summary variable.  
Recent work studies measures of how well-suited ignorability is for a given analysis, and devises a sensitivity analysis for scenarios where the ignorable and stable proxy assumptions may not hold \citep{little2020measures}. 

In the distribution shift literature, the scaled likelihood estimator is frequently deployed in label shift settings.  
For example, following the guidance of \citet{alexandari2020maximum}, \citet{sun2022beyond} develop a method to adapt predictions at test time when potentially spurious correlates can be measured. 

Missing data estimators are also frequently used to analyze health  data.  Multilevel regression and post-stratification \citep{wang2015forecasting} --- a relative of ignorable estimators --- have been applied to health survey data to estimate, for example, disease prevalence at fine levels of spatial resolution \citep{zhang2015validation}.





\vspace{-1em}
\section{Experiments}

\vspace{-0.5em}
We study our proposed stable proxy estimation procedure alongside typical ignorable estimators under a variety of conditions (e.g., well-specified and misspecified) in a simulation study (Section \ref{sec:simulation-study}), a semi-synthetic study (Section~\ref{sec:semi-synthetic}) and a case study using a large health survey (Section \ref{sec:case-study}).
In each setting, we compare a variety of calibrated and uncalibrated estimators for $\mu_Y^{(1)}$, including:
\vspace{-.5em}
\begin{itemize}\setlength\itemsep{-.2em}
    \item \texttt{cc}: complete case estimate, i.e., $\hat{\mu}_Y^{(0)}$;
    \item \texttt{ipw} (\texttt{cipw}): (calibrated) inverse probability weighted estimator, using the propensity score model $P(M \,|\, X)$,  assuming ignorability\footnote{In practice, importance sampling-based estimators such as IPW tend to have high variance, so we use normalized weights \citep{hirano2003efficient, robins2000marginal}, and in the case study also use clipping or truncation of the largest weights to further reduce variance \citep{ionides2008truncated}.};
    \item \texttt{direct} (\texttt{cdirect}): (calibrated) direct prediction using $P(Y \,|\, X, M=0)$, assuming ignorability;
    \item \texttt{proxy} (\texttt{cproxy}): (calibrated) auxiliary proxy variable estimator, assuming stable proxies.
\end{itemize}
\vspace{-.5em}
For each estimate, we first fit a propensity score model if needed, i.e., $p(M \,|\, X)$ and a prediction score model, i.e., $p(Y \,|\, X, M=0)$, using half of the data, tuning hyperparameters with internal ten-fold cross validation.  
We use the other half as a validation set to first potentially calibrate and then form the final estimate of $\mu_Y^{(1)}$. 
For propensity scores, we use xgboost \citep{chen2016xgboost}, and for prediction scores we compare using xgboost and naive Bayes (modeling each dimension of $X$ as a conditionally independent Bernoulli).


\subsection{Simulation study}
\label{sec:simulation-study}

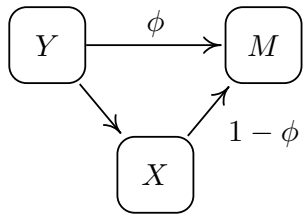
\begin{figure}
\begin{center}
\scalebox{1.15}{
\begin{tikzpicture}[>=stealth',shorten >=1pt,auto,node distance=3cm, semithick]
  \tikzstyle{every state}=[fill=none,shape=rectangle,rounded corners=2mm,text=black]
  \draw (-2.5,1.5) node[state, fill=white] (Y) {$Y$};
  \draw (0,1.5) node[state, fill=white] (M) {$M$};
  \draw (-1.25,0) node[state] (X) {$X$};
  \draw [-arcsq] (Y) -- (X);
  \draw [-arcsq] (Y) -- (M);
  \draw [-arcsq] (X) -- (M);
  \draw (-1.25,1.75) node[] (l) {$\phi$};
  \draw (0,.48) node[] (l) {$1-\phi$};
\end{tikzpicture}
}
\end{center}
\caption{Simulation study setup. Data are generated from the above graph, where scalar parameter $\phi$ trades off between ignorable (i.e., $\phi=0$) vs.~stable proxy (i.e., $\phi=1$) missing data mechanisms for $M$.}
\label{fig:simulation-graph}
\end{figure}

\vspace{-0.6em}
We study the estimator in a controlled setting, generating synthetic data according to the graph depicted in Figure~\ref{fig:simulation-graph}.  The full data generating procedure is
\begin{align*}
    Y &\sim \texttt{bern}(\mu_Y) \\
    X_d \,|\, Y=y &\sim \texttt{bern}\left(m_{d,y}\right) \text{ for } d=1, \dots, D \\
    M \,|\, X=x, Y=y &\sim \texttt{bern}\left(\sigma(g_\phi(x,y))\right)
\end{align*}
where 
\begin{align}
g_\phi(x, y) &= \beta_0 + \beta \cdot \left((1-\phi)\cdot \bar{x} + \phi \cdot y)\right) \, ,
\label{eq:missing-mechanism}
\end{align}
$\sigma$ is the inverse logit link, and $\bar{x}$ is the average of the $D$ elements of $x$.  
$\phi$ controls the degree to which the missing indicator $M$ depends on $Y$ or $X$ (or both).
We examine a grid of $\phi$ values, interpolating between ignorability ($\phi = 0$) and stable proxies ($\phi = 1$). 
When $0 < \phi < 1$ all estimators are misspecified.
We fix the complete data mean $\mu_Y = .35$, and choose $\beta_0$ and $\beta$ so that $\mu_Y^{(0)} = .25$ and $\mu_Y^{(1)} = .5$ throughout.

\vspace{-.5em}
\paragraph{Toy simulation}
We first examine estimators in a simpler low-dimensional setting.
We let $X$ be a binary random variable, with $m_{1,y=0} = .3$ and $m_{1,y=1}= .7$.

\begin{figure}[t!]
    \centering
     \begin{subfigure}[b]{0.45\textwidth}
         \centering
         \includegraphics[width=\textwidth]{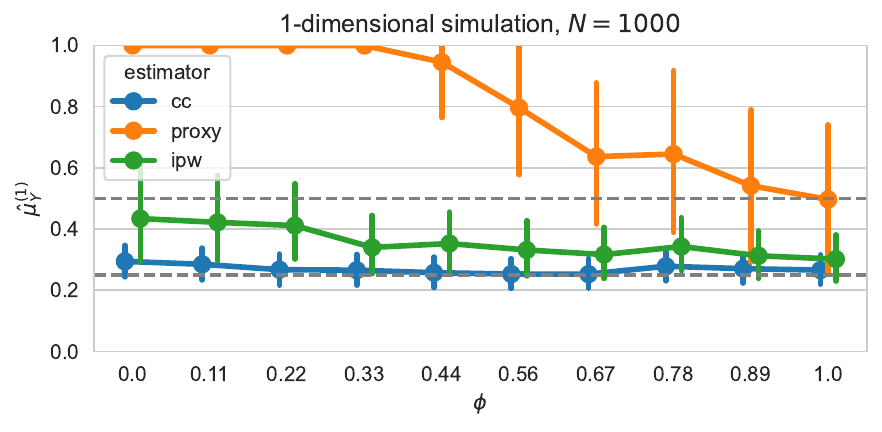}
         \caption{Toy estimates, $N=1{,}000$}
         \label{fig:toy_estimates}
     \end{subfigure}
     
     \begin{subfigure}[b]{0.45\textwidth}
         \centering
         \includegraphics[width=\textwidth]{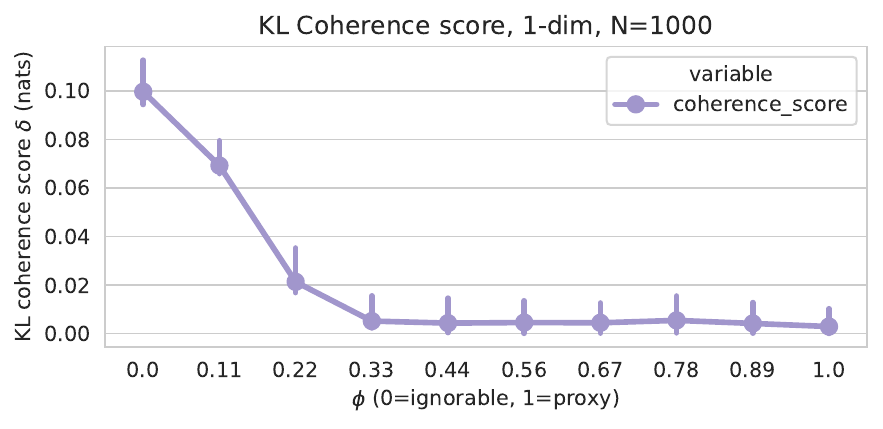}
         \caption{Toy coherence scores, $N=1{,}000$}
         \label{fig:toy_KL}
     \end{subfigure}
    \begin{subfigure}[b]{0.45\textwidth}
         \centering
         \includegraphics[width=\textwidth]{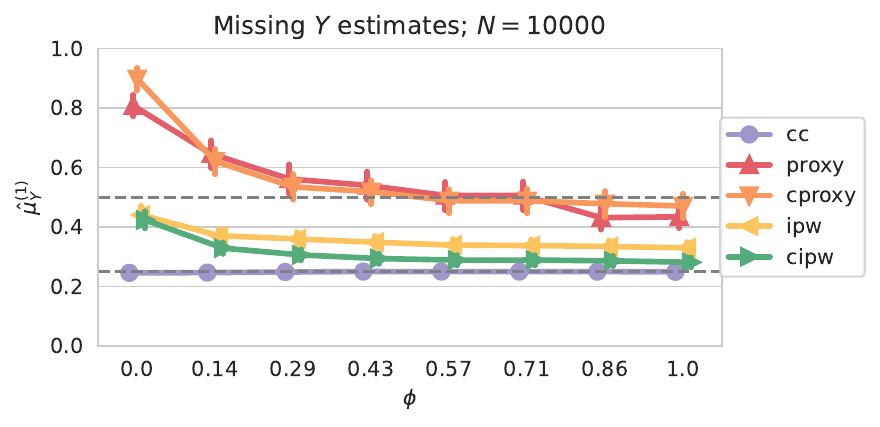}
         \caption{High-dim estimates, $N=10{,}000$}
         \label{fig:highdim_estimates}
    \end{subfigure}
    
    \begin{subfigure}[b]{0.45\textwidth}
         \centering
         \includegraphics[width=\textwidth]{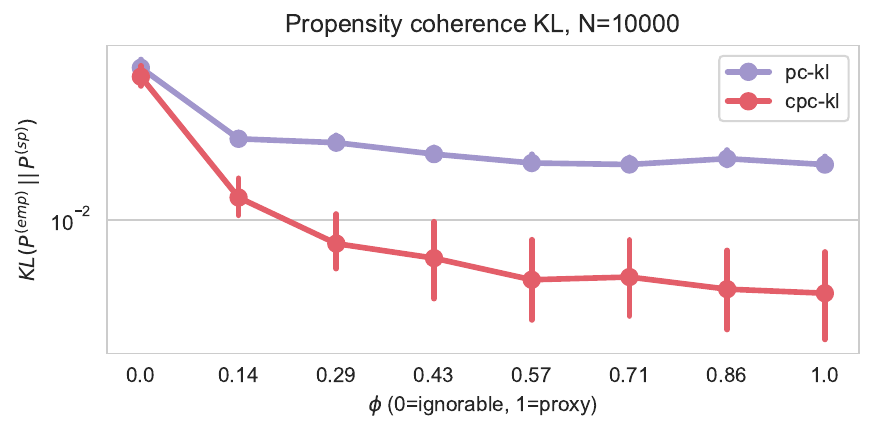}
         \caption{High-dim coherence scores, $N=10{,}000$}
         \label{fig:highdim_estimates}
    \end{subfigure}

    \caption{Simulation study results. (a) and (c) compare estimator performance in the toy and high-dimensional simulation studies as a function of $\phi$ (interpolating between ignorability at $\phi=0$ and stable proxies at $\phi=1$), with 95\% bootstrap CIs. (b) and (d) show coherence scores as a function of $\phi$. 
    }
    \label{fig:simulation-study}
    \vspace{-.5em}
\end{figure}

\begin{figure*}[t!]
    \centering
     
     \begin{subfigure}[b]{\textwidth}
         \centering
         \includegraphics[width=.48\textwidth]{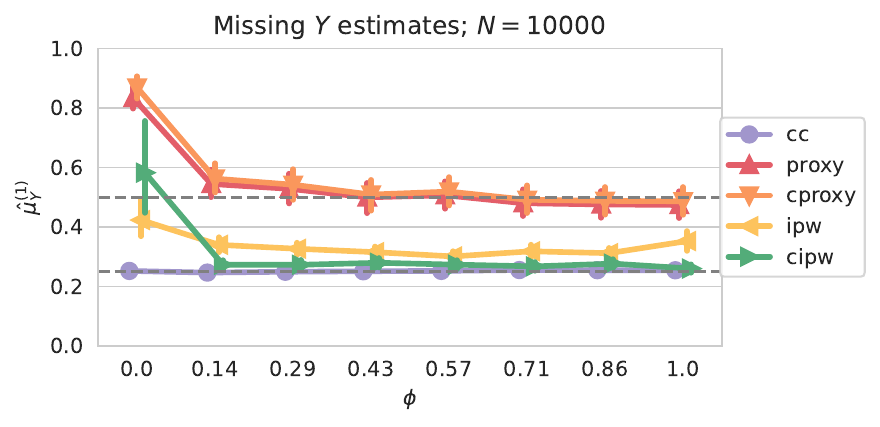}%
         \includegraphics[width=.45\textwidth]{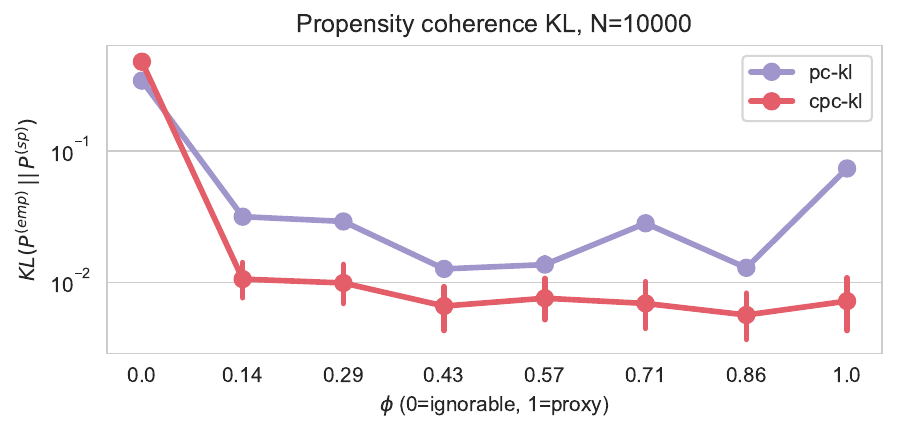}
         \caption{xgboost, $N=10{,}000$}
         \label{fig:semi-synthetic-xgboost-10k}
     \end{subfigure}

    \begin{subfigure}[b]{\textwidth}
         \centering
         \includegraphics[width=.48\textwidth]{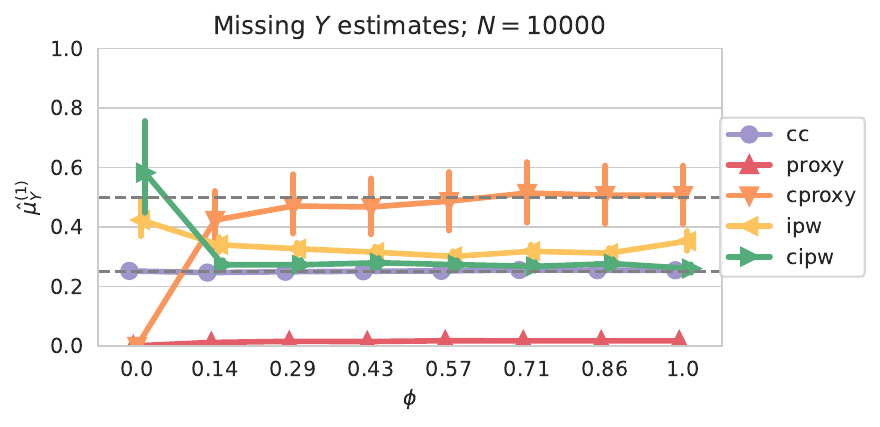}%
         \includegraphics[width=.45\textwidth]{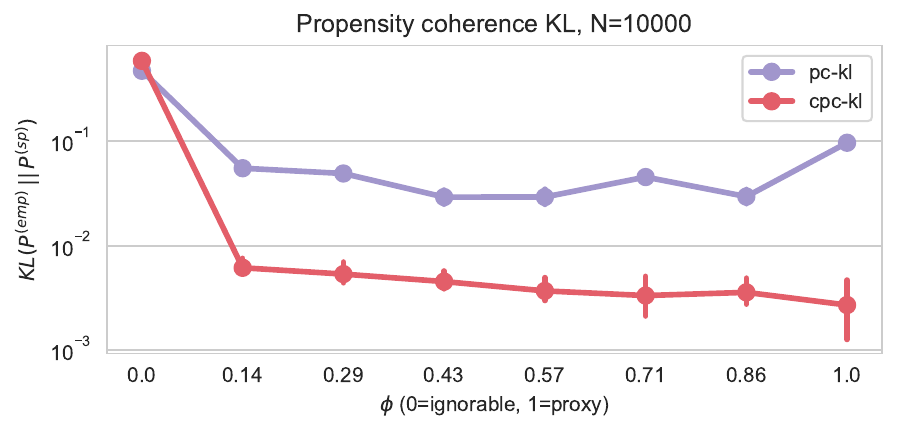}
         \caption{Naive Bayes, $N=10{,}000$}
         \label{fig:semi-synthetic-nb}
     \end{subfigure}
     

    \caption{Semi-synthetic results.  
    The missing data mean is 0.5 and observed data mean is 0.2). 
    (a) shows results where $P(Y \,|\, X)$ was estimated with xgboost, with estimator performance on the left and coherence scores on the right. 
    (b) shows results where $P(Y \,|\, X)$ was instead estimated with naive Bayes.
    The un-calibrated \texttt{proxy} estimator from the less flexible naive Bayes model makes extreme and incorrect estimates that are overly confident.  
    After calibration, the \texttt{cproxy} estimator makes correct predictions as the model becomes better specified (as $\phi$ increases).  
    Estimators based on the more accurate xgboost model do not suffer from this pathology.
    }
    \label{fig:semi-synthetic}
    \vspace{-1em}
\end{figure*}



\vspace{-.5em}
\paragraph{High-dimensional simulation}
We extend the toy simulation by simulating a 100-dimensional $X$ vector, where each $m_{d,y=0} = .42$ and $m_{d,y=1} = .5$.


\vspace{-.5em}
\paragraph{Results}
Figure~\ref{fig:simulation-study} shows the main results. In both the toy and high-dimensional settings, we observe the expected tradeoff: when $\phi=0$, the ignorable estimators, \texttt{ipw} and \texttt{cipw}, have the lowest error in estimating the true $\mu_Y^{(1)} = 0.5$, while the non-ignorable stable proxy estimators perform poorly. 
Likewise, when $\phi=1$, the stable proxy estimators are most accurate, while the incorrect ignorability assumption causes extreme bias for \texttt{ipw} and \texttt{cipw}. 
In the middle regime where $0 < \phi < 1$, both methods have invalid assumptions. 
However, for values of $\phi$ near 0 the ignorable estimators tend to do better, while near 1 the stable proxy estimators do better. 
We also see that the \texttt{proxy} estimator in the toy setting is extremely sensitive to misspecification, while in the higher-dimensional setting it is more robust. 
We also observe the expected behavior from the propensity coherence score --- as $\phi$ increases, the divergence between propensity score models shrinks to zero. 

Calibration did not seem to matter in the toy setting, so calibrated estimators were not shown for clarity. 
In the high-dimensional setting, the importance of calibration is stark. 
The calibrated estimators dominate the uncalibrated ones, and the coherence scores using calibrated propensities are also always lower. 
Interestingly, the effect of calibration holds despite the large $N$ setting of the high-dimensional problem. 
\begin{figure*}[t!]
    \centering
     \begin{subfigure}[b]{\textwidth}
         \centering
         \includegraphics[width=.4\textwidth]{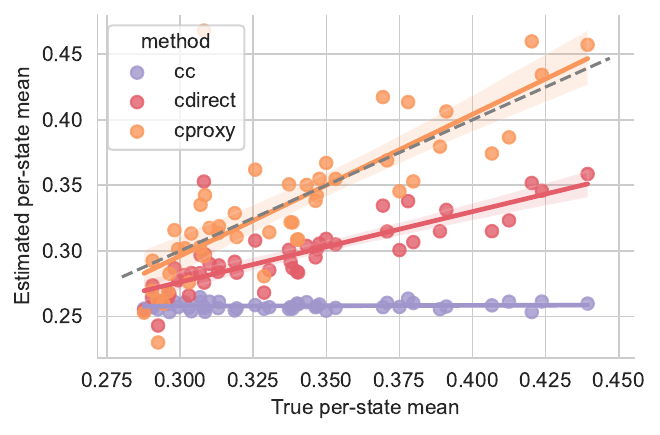} %
         \includegraphics[width=.4\textwidth]{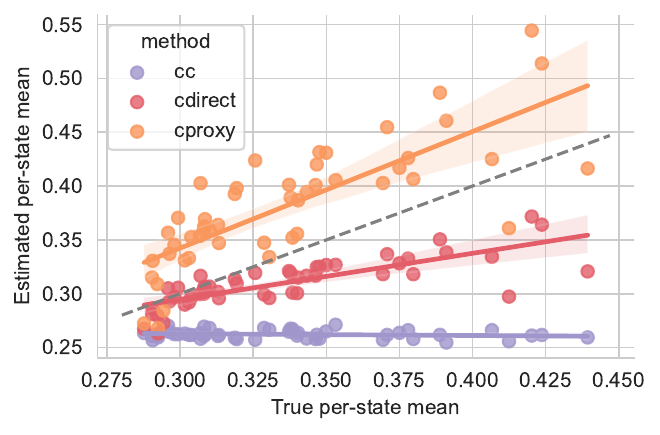}
         \caption{Blood pressure prevalence estimates.  (left) Colorado as source state, (right) Utah as source state.}
         \label{fig:three sin x}
     \end{subfigure}
     \begin{subfigure}[b]{\textwidth}
         \centering
         \vspace{1em}
         \includegraphics[width=.4\textwidth]{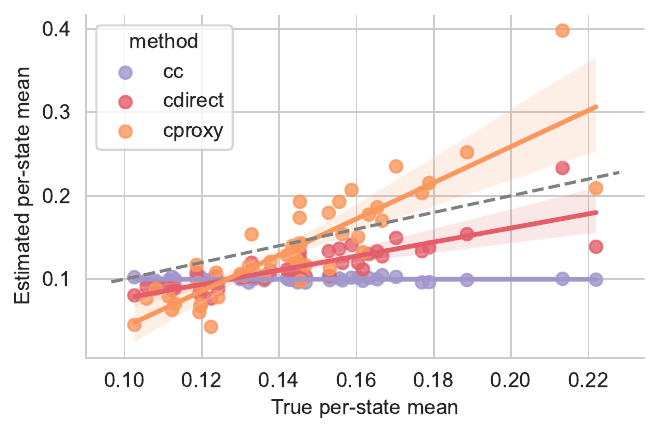} %
         \includegraphics[width=.4\textwidth]{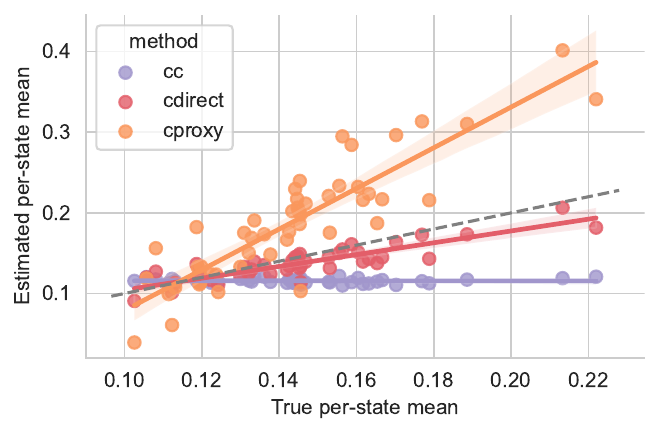}
         \caption{Diabetes prevalence estimates.  (left) Colorado as source state, (right) Utah as source state.}
         \label{fig:five over x}
     \end{subfigure}


    \caption{Case study results.  Each dot is a target state, with an estimate from the source state (Colorado or Utah) on the $Y$-axis, true value on the $X$-axis (with per-method linear fit), and $Y=X$ line in grey.  (a) Blood pressure, (b) Diabetes.
    }
    \label{fig:case-study}
    \vspace{-.5em}
\end{figure*}

\subsection{Semi-synthetic data --- health survey}
\label{sec:semi-synthetic}

\vspace{-0.5em}
We use data from the Behavioral Risk Factor Surveillance System (BRFSS) \citep{brfss}, a health survey run by the CDC.  
The BRFSS is a telephone survey that collects data from U.S. residents about health-related risk behaviors, chronic health conditions, and use of preventative services.
We use these survey data to estimate the prevalence of a disease using other proxy variables available in the survey.  
We first resample the dataset (with replacement) according to the survey weights of each unit provided by the BRFSS study to make the dataset broadly representative of each state, and avoid using weighted estimators. 
We then assemble a 107-dimensional binary feature vector derived from demographic information and other conditions.  
In this section, we take $Y$ to indicate high blood pressure.
We refer to Appendix~\ref{sec:app-results} for details on dataset processing.

Similar to the simulation studies, we induce missingness as a function of the covariates $X$ and outcome $Y$ (Eq.~\ref{eq:missing-mechanism}), varying $\phi$ from 0 (ignorable) to 1 (stable proxy).  
A sampling of results are displayed in Figure~\ref{fig:semi-synthetic}.
We observe similar behavior ---  that the stable proxy estimator dominates as $\phi$ approaches one --- now on real health survey data using an estimate of $P(Y \,|\, X)$. 

These results also highlight the importance of both prediction model quality and calibration. 
The uncalibrated naive Bayes model forms estimates that are pathologically bad (Figure~\ref{fig:semi-synthetic-nb}), whereas the uncalibrated xgboost model produces good estimates at the same sample size (Figure~\ref{fig:semi-synthetic-xgboost-10k}).
Even when the model is well-calibrated, the resulting error bars are substantially wider for the worse naive Bayes model.  
Additionally, we observe that the propensity coherence score closely tracks the quality of the estimate, but only using well-calibrated probabilities.  
In fact, using uncalibrated probabilities to compute the propensity coherence score can result in non-monotonically decreasing behavior as $\phi$ approaches one --- a property that vanishes when using calibrated scores. 



\subsection{Case study: natural missingness} 
\label{sec:case-study}

\vspace{-0.5em}
Lastly, we study a set of more natural missingness patterns in BRFSS.
We consider the estimation of the prevalence of a chronic disease, where the missing data correspond to a target state and the observed data correspond to a source state, using the same set of covariates $X$ mentioned in the previous section and described in full in Appendix~\ref{sec:app-results}.
We study two outcomes, high blood pressure and diabetes, with a third (depression) and fourth (high cholesterol) in Appendix~\ref{sec:app-results}. 

We use two source states, Colorado and Utah, to extrapolate estimates of the outcome of interest to the other states. 
We select Colorado and Utah for their relatively low rates of high blood pressure and diabetes, making the difference between the observed and missing mean more extreme.  
We show high blood pressure and diabetes results in Figure~\ref{fig:case-study}, and depression and high cholesterol results in Figure~\ref{fig:case-study-app} of Appendix~\ref{sec:app-results}. 
Figure~\ref{fig:case-study-results-table} of Appendix~\ref{sec:app-results} contains a table of additional quantitative results, reporting aggregate mean absolute percent errors for the performance of each estimator on each task.

No estimator dominates in all settings.  
When estimating high blood pressure prevalence using Colorado as the source, the stable proxy estimator outperforms the ignorable estimator.
However, using Utah as the source, we see nearly the correct slope but a consistent bias, suggesting the stable proxy assumption may be less appropriate.
For diabetes prevalence estimation, the ignorable estimator performs better for both Colorado and Utah as source states, suggesting that ignorability is more appropriate in this setting. 
Our results suggest that neither ignorability nor stable proxies are universally valid assumptions, and that the true missingness structure likely lies somewhere in between.  



\vspace{-1em}
\section{Discussion}
In this work, we formalized a connection between non-ignorable missing data estimation and the label shift problem.  Our empirical study showed that the scaled maximum likelihood estimator can be effective in the non-ignorable setting, in synthetic, semi-synthetic, and real health survey data. 
We also developed the propensity coherence score to measure how concordant the stable proxy assumption is with observed data, and showed it tracks closely with estimate quality.
In future work, we aim to develop more practical guidance for using the coherence score, and will attempt to incorporate the score into the estimate itself, perhaps by inflating uncertainty or correcting for bias.
Another potential use case is decomposing generalization error in new populations due to misspecified assumptions, perhaps in the style of \citet{cai2023diagnosing}, which disentangles error attributable to shifts in $X$ and instabilities in $P(Y \,|\, X)$.

We hope this work encourages the distribution shift community to examine the tools developed for missing data --- e.g., sensitivity analyses and propensity scores --- to cope with the reality that often neither ignorability nor the label shift assumption hold.  






\subsubsection*{Acknowledgements}
The authors would like to thank Antoine Wehenkel and Sean Jewell for feedback on an earlier version of this work.

\bibliographystyle{apalike}
\bibliography{refs.bib}


\clearpage
\appendix

\thispagestyle{empty}

\onecolumn{\makesupplementtitle}


\section{Propensity Coherence Score Derivation}
\label{sec:consistency}

Given the missing data pattern and observed covariates, it is straightforward to estimate a ``propensity score'' 
\begin{align}
    e(x) &\triangleq p(m=1 \,|\, x) \,,
\end{align}
which is typically used to adjust estimates in the ignorable setting.  
However, even if such adjustments are insufficient and a different assumption such as label shift (or stable proxies) is more appropriate, we can still make use of propensity scores for a type of sensitivity analysis.
Given the label shift assumption, i.e., $p(x \,|\, y, m=0) = p(x \,|\, y, m=1)$, we can form an estimate of $p(m = 1 \,|\, x)$ \emph{indirectly}, using the scores from our predictive model $p(y \,|\, x)$, our estimand $\mu_Y^{(1)}$, the observed data mean $\mu_Y^{(0)}$, and the prevalence $p(m)$.  



If the label shift assumption is true (and our estimators are consistent, i.e. are well specified and there is low finite sample estimation error), these estimates should match.  
Mathematically, we will show that they should exactly match given only the label shift assumption.
This ``stress test'' allows us to inspect the validity of the label shift assumption, supposing only that we were able to form a well-calibrated estimate of $p(m \,|\, x)$ well --- a task no more difficult a task than estimating $p(y \,|\, x)$ in the binary setting. 





We can write $p(m \,|\, x)$ as:
\begin{align}
    p(m \,|\, x) &\propto p(x \,|\, m) p(m) = \left(\sum_{y \in \{0,1\}} p(x, y \,|\, m) 
    \right)p(m).
\end{align}

Conditioned on $m$, the joint of $x$ and $y$ can be rewritten as:
\begin{align}
    p(x, y \,|\, m) &= p(x \,|\, y, m) p(y \,|\, m) \\
    &= p(x \,|\, y, m=0) p(y \,|\, m) && \text{ label shift assumption } \\
    &\propto \frac{p(y \,|\, x, m=0)}{p(y \,|\, m=0)} p(y \,|\, m). && \text{ scaled likelihood (i.e., Bayes) }
\end{align}

Plugging this back in, we get :
\begin{align}
    p(m \,|\, x)
    &\propto \left(\sum_{y \in \{0,1\}} p(x, y \,|\, m) 
    \right)p(m) \\
    &\propto \left(\sum_{y \in \{0,1\}} \frac{p(y \,|\, x, m=0)}{p(y \,|\, m=0)} p(y \,|\, m) \right) p(m).
\end{align}
All that remains are terms that are estimated already during our procedure to estimate the mean of the missing $Y$'s .  
So we can compute what $p(m \,|\, x)$ is implied by the label shift assumption by just normalizing the above across both levels of $m$.  

For binary $y$ and $m$, we can construct an estimate of $p(m \,|\, x)$ by plugging in estimators for each component.  Given a predictive model fit on the observed $Y$'s where $M=0$, we have that $f(x) \approx p(y \,|\, x, m=0)$, $p(y \,|\, m=0)$ is just the mean of the observed $Y$'s so we can use the sample mean, $p(y \,|\, m=1)$ is our estimand of interest in the first place, and $p(m)$ can also be estimated using e.g. the sample proportion of $Y$'s that are missing.




\section{Additional Results and BRFSS Details}
\label{sec:app-results}

\paragraph{BRFSS processing}
For predicting blood pressure, the features used include:

\texttt{['sex',
 'race',
 'smoker',
 'smoked\_100\_cigs',
 'income',
 'marital',
 'education',
 'drinks',
 'binge\_drink',
 'lives\_metro',
 'gen\_health',
 'days\_bad\_phys\_health',
 'days\_bad\_mental\_health',
 'insurance',
 'have\_pcp',
 'last\_checkup',
 'exercise',
 'rent\_own\_house',
 'military',
 'employment',
 'has\_kids',
 'deaf',
 'blind',
 'difficulty\_stairs',
 'cholesterol',
 'heartattack',
 'cad',
 'stroke',
 'asthma',
 'skin\_cancer',
 'other\_cancer',
 'copd',
 'depression',
 'kidney',
 'diabetes',
 'arthritis',
 'age\_bin',
 'height\_bin',
 'weight\_bin',
 'bmi\_bin'].
}

We use similar features for the other outcomes, \texttt{diabetes}, \texttt{depression}, and \texttt{cholesterol}, removing them from the set of features when they are used as the outcome. 

All features are mapped to a one-hot encoding (removing the first entry) and then concatenated.
Unknown or missing features are coded as a separate catgegory, with the exception of height and weight, which are first imputed via multiple imputation (with the \texttt{mice} package in R).
For each condition of interest that we wish to predict as the label $Y$, this results in a binary vector of size 107. 
The dataset itself is publicly available via the CDC.
We will release the minimal preprocessing code along with the code to reproduce all experiments and figures with the camera ready.

\paragraph{Additional BRFSS case study results}
Figure 6 below shows additional results for estimating depression prevalence and high cholesterol prevalence, again using Colorado and Utah as source states. 

Lastly, Figure 7 shows a table of mean absolute errors for different estimators for the 4 prevalences of interest, for Colorado and Utah as source states. 95\% CIs from bootstrapping are also shown.


\begin{figure*}[h]
    \centering


      \begin{subfigure}[b]{0.48\textwidth}
         \centering
         \includegraphics[width=\textwidth]{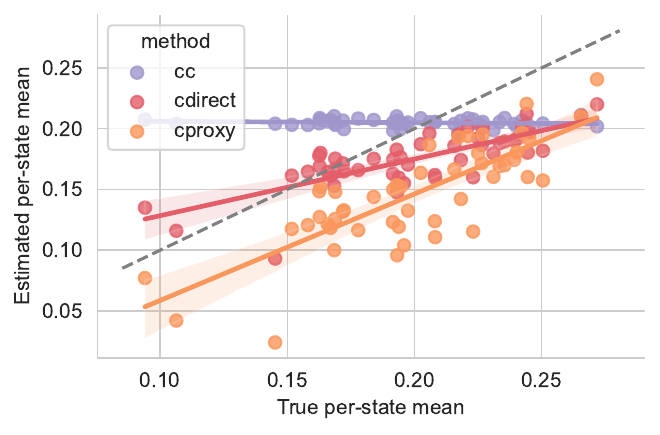}
         \caption{Depression prevalence estimates (Colorado as source)}
         \label{fig:three sin x}
     \end{subfigure}
     \hfill
     \begin{subfigure}[b]{0.48\textwidth}
         \centering
         \includegraphics[width=\textwidth]{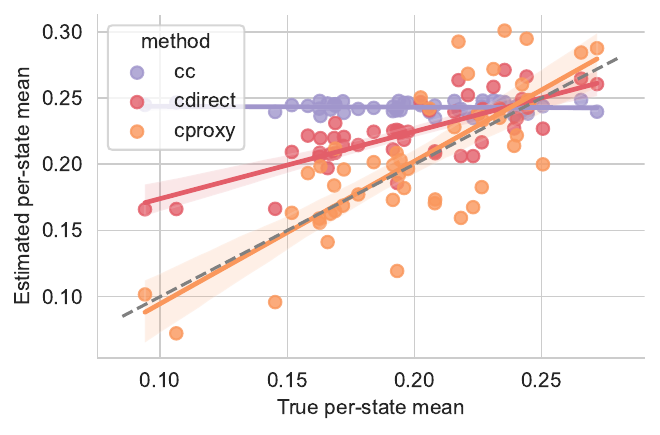}
         \caption{Depression prevalence estimates (Utah as source)}
         \label{fig:five over x}
     \end{subfigure}

     \begin{subfigure}[b]{0.48\textwidth}
         \centering
         \includegraphics[width=\textwidth]{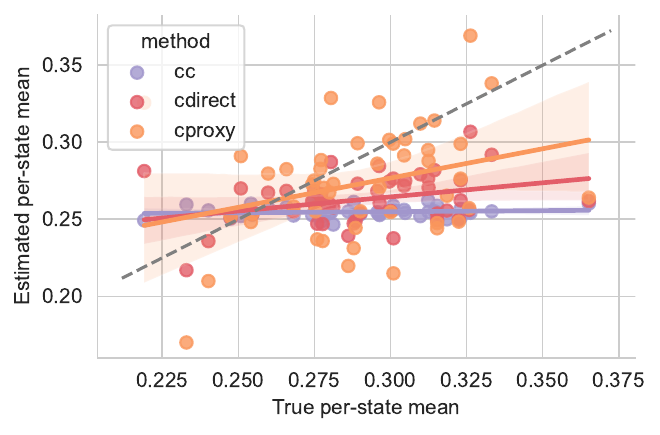}
         \caption{High cholesterol prevalence estimates (Colorado as source)}
         \label{fig:three sin x}
     \end{subfigure}
     \hfill
     \begin{subfigure}[b]{0.48\textwidth}
         \centering
         \includegraphics[width=\textwidth]{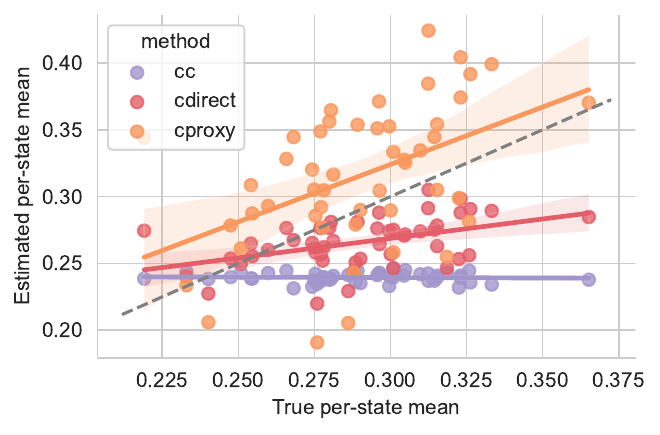}
         \caption{High cholesterol prevalence estimates (Utah as source)}
         \label{fig:five over x}
     \end{subfigure}
    \caption{Additional case study summaries, now for the tasks of estimating depression and high cholesterol prevalence again using Colorado and Utah as source states.}
    \label{fig:case-study-app}
\end{figure*}

\begin{figure*}[h]

     \centering
     \scalebox{.9}{\begin{tabular}{llllll}
\toprule
   & method &                 cc &            cdirect &               cipw &             cproxy \\
{} & {} &                    &                    &                    &                    \\
\midrule
CO & bloodpressure &  0.223 [0.22-0.23] &  0.122 [0.12-0.13] &  0.534 [0.43-0.62] &  0.073 [0.06-0.08] \\
   & diabetes &  0.285 [0.28-0.29] &  0.213 [0.21-0.22] &  0.986 [0.73-1.22] &  0.234 [0.22-0.25] \\
   & cholesterol &  0.128 [0.12-0.13] &  0.106 [0.10-0.11] &  0.577 [0.45-0.70] &  0.114 [0.10-0.13] \\
   & depression &  0.182 [0.18-0.19] &  0.143 [0.13-0.15] &  0.741 [0.62-0.87] &  0.278 [0.26-0.29] \\
UT & bloodpressure &  0.210 [0.20-0.21] &  0.076 [0.07-0.08] &  0.344 [0.28-0.40] &  0.152 [0.14-0.16] \\
   & diabetes &  0.190 [0.18-0.20] &  0.071 [0.07-0.08] &  0.513 [0.41-0.62] &  0.364 [0.34-0.39] \\
   & cholesterol &  0.172 [0.17-0.18] &  0.095 [0.09-0.10] &  0.429 [0.36-0.51] &  0.159 [0.15-0.17] \\
   & depression &  0.285 [0.28-0.29] &  0.170 [0.16-0.18] &  0.634 [0.54-0.73] &  0.140 [0.12-0.16] \\
\bottomrule
\end{tabular}
}
     \caption{Mean absolute error with 95\% CIs from bootstrap from the BRFSS case study, for each method, each source state (Colorado and Utah) and each outcome. \texttt{cipw} is shown in this table but not in the summary results figures that show results from each target state, since its overall performance was so much worse.  }
     \label{fig:case-study-results-table}
\end{figure*}


\end{document}